\documentclass{article}

 \usepackage[final, nonatbib]{neurips_2025_ml4ps}

\usepackage[utf8]{inputenc} % allow utf-8 input
\usepackage[T1]{fontenc}    % use 8-bit T1 fonts
\usepackage{hyperref}       % hyperlinks
\usepackage{url}            % simple URL typesetting
\usepackage{booktabs}       % professional-quality tables
\usepackage{amsfonts}       % blackboard math symbols
\usepackage{nicefrac}       % compact symbols for 1/2, etc.
\usepackage{microtype}      % microtypography
\usepackage{xcolor}         % colors
\usepackage{amsmath}
\usepackage{graphicx}

\title{Knowledge is Overrated: A zero-knowledge machine learning and cryptographic hashing-based framework for verifiable, low latency inference at the LHC}
\author{%
  Pratik Jawahar \\
  University of Manchester \\
  \texttt{pratik.jawahar@cern.ch} \\
  \And
  Caterina Doglioni \\
  University of Manchester \\
  \texttt{caterina.doglioni@cern.ch} \\
  \AND
  Maurizio Pierini \\
  CERN \\
  \texttt{maurizio.pierini@cern.ch}
}

\begin{document}

\maketitle

\begin{abstract}
  Low latency event-selection (trigger) algorithms are essential components of Large Hadron Collider (LHC) operation. Modern machine learning (ML) models have shown great offline performance as classifiers and could improve trigger performance, thereby improving downstream physics analyses. However, inference on such large models does not satisfy the $40\text{MHz}$ online latency constraint at the LHC. In this work, we propose \texttt{PHAZE}~\cite{pratik_jawahar_2025_17370252}, a novel framework built on cryptographic techniques like hashing and zero-knowledge machine learning (zkML) to achieve low latency inference, via a certifiable, early-exit mechanism from an arbitrarily large baseline model. We lay the foundations for such a framework to achieve nanosecond-order latency and discuss its inherent advantages, such as built-in anomaly detection, within the scope of LHC triggers, as well as its potential to enable a dynamic low-level trigger in the future.
\end{abstract}

\section{Introduction}
\label{sec:intro}
Accelerating ML inference to satisfy the LHC's operational constraints is currently approached as an ad-hoc optimization problem driven by an accuracy-acceleration tradeoff. Algorithms like AXOL1TL and CICADA, recently tested~\cite{Gandrakota:2024yqs} by the CMS experiment, combine this type of an approach with hardware-specific optimization techniques, using frameworks like \texttt{hls4ml}~\cite{fahim2021hls4ml}. These algorithms successfully achieve $O(\mathrm{ns})$ latencies for anomaly detection triggers~\cite{Gandrakota:2024yqs}. However, we hypothesize that, to achieve $O(\mathrm{ns})$ latencies for a larger set of ML-based trigger decisions, we need a novel approach to trigger design compatible with dynamic performance-improvement strategies.

We start with a short primer on polynomial fingerprinting, zkML, and early-exiting (EE) that form the building blocks of our framework.

\subsection{Polynomial fingerprinting}

\paragraph{Hash function:}
Any function that maps a mathematical object, like a vector of arbitrary size, to a fixed-size value~\cite{preneel1994cryptographic, rogaway2004cryptographic}. We restrict the definition to non-cryptographic hash functions~\cite{estebanez2014performance} since we do not require security-related properties in the hashing algorithm for our framework. Say we have a family of hash functions $\mathcal{H}\{h: \mathcal{M}\rightarrow \mathcal{Y} \}$. We can then define a probabilistic hashing algorithm as a scheme, that for a given arbitrary-size object $M_i \in \mathcal{M}$, chooses a random hash function $h_i \in \mathcal{H}$ to evaluate the hash.

For a probabilistic hash to serve as a near-unique fingerprint for all objects in $\mathcal{M}$ we require that the collision probability~\cite{rogaway2004cryptographic} over a random $h_i$, for any $2$ arbitrary values $x_1, x_x \in \mathcal{M}$, be bounded by a small, well-defined value $\epsilon$, written as,

\begin{equation}
    \Pr_{h_i\in \mathcal{H}} [h_i(x_1) = h_i(x_2)] \leq \epsilon
\end{equation}

\paragraph{Rabin fingerprinting~\cite{rabin1981fingerprinting}:}
A hash defined over a finite, Galois field, $\mathrm{GF(2)}$, that maps an $n$-bit vector represented as a polynomial (using techniques like polynomial interpolation~\cite{bleichenbacher2000noisy}), $P(x)$ of degree $(d-1)$, to the remainder, $r(x)$, obtained on dividing $P(x)$ by $g(x)$; where $g(x)$ is an irreducible polynomial of degree $k$ over $\mathrm{GF(2)}$\cite{rabin1981fingerprinting}. Thereby the Rabin Fingerprinting algorithm reduces a large $d$-bit vector to a unique $k$-bit fingerprint.

So the Rabin fingerprint over a large finite field of prime order, $\mathbb{F}_p$, is given by,

\begin{equation}
    h(x) = P(x)(\bmod~g(x))
    \label{eq:rabinfingerprint}
\end{equation}

where the coefficients of $h(x)$ (degree $k-1$) are the reduced $k{\text -}\mathrm{bit}$ hash.

The collision probability is then described by the Schwartz-Zippel Lemma~\cite{ore1921höhere}. The lemma states that for any two distinct polynomials $P_1(x), P_2(x)$, each of degree at most $(d-1)$, the collision probability for a randomly chosen point $q \in \mathbb{F}_p$ is bounded by,

\begin{equation}
  \Pr_{q\in \mathbb{F}_p} [P_1(q) = P_2(q)] \leq \frac{(d-1)}{|\mathbb{F}_p|}  
  \label{eq:schwartz-zippel}
\end{equation}

Thereby, choosing a finite field ordered by a large enough prime, and constraining the highest polynomial degree to be low, leads to statistically negligible collision probabilities.

\subsection{Zero-knowledge machine learning}
\label{sec:zkML}
\paragraph{Zero-Knowledge Proof:}
A ZKP is a cryptographic protocol that allows one party, the prover, to convince another party, the verifier, that a given statement is true, without revealing any information beyond the fact of the statement's truth~\cite{goldreich1996composition}. A powerful application of modern ZKP systems lies in their ability to guarantee computational integrity, making them tools for fast verifiability. The core properties of ZKPs are completeness, soundness and privacy-preservation.

\paragraph{zkML:}
An application of ZKPs for generating a succinct proof, $\pi$, of the correct execution of an ML model's inference computation, $C(x,w)=y$, where $x\in L$ is a public input in a corpus $L$ (e.g., training dataset), $w$ is the private witness (e.g., model parameters, intermediate activations) and $y$ is the public model output. For our usage, privacy is not a major factor and we focus on verifiability, which for a given prover $\mathcal{P}$ and verifier $\mathcal{V}$ is defined, as in~\cite{ben2015secure}, by the completeness,

\begin{equation}
    \Pr_{}[\langle \mathcal{P}(x,w) \leftrightarrow \mathcal{V}(x)\rangle = \mathrm{accept}] = 1
\end{equation}

and the soundness, for $x'\notin L$ and a random prover $\mathcal{P}^*$, 
\begin{equation}
    \Pr_{}[\langle \mathcal{P}^*(x',w) \leftrightarrow \mathcal{V}(x')\rangle = \mathrm{accept}] \leq \epsilon
\end{equation}

where $\langle\rangle$ represents prover-verifier interactions.

The foundational step of most zkML systems is arithmetization; the process of transforming computation $C$ into a family of algebraic equations. These equations serve as polynomial constraints that are tractable when defined over a finite field $\mathbb{F}_p$, and the validity of the computation is reduced to proving that these constraints hold. The complexity of a zkML system is largely concentrated in the arithmetization stage. The subsequent steps involve using cryptographic primitives such as hashing and Polynomial Commitment Schemes (PCS)~\cite{goldreich1996composition} to compress the constraints down to unique, tractable commitments (e.g., fingerprints), used for faster verification. The Schwartz-Zippel lemma (Eq.\ref{eq:schwartz-zippel}) again serves as a bound for the integrity of the proof.

We focus on a specific type of zkML system based on zk-STARK (Zero-Knowledge Scalable Transparent ARguments of Knowledge)~\cite{ben2018scalable}. zk-STARKs are typically constructed under minimal cryptographic assumptions which gives rise to scalability and transparency. The complexity, for a given computation of size $T(n)$, scales quasi-linearly, $O(T(n)\mathrm{poly}\log T(n))$ for proof generation and poly-logarithmically, $O(\mathrm{poly}\log T(n))$ for proof verification~\cite{ben2018scalable}. The transparency makes the system auditable and reproducible.

\subsection{Early-Exit}
\label{sec:ee}
EE~\cite{teerapittayanon2016branchynet} strategies are a popular approach for significantly reducing inference time and computation of AI algorithms running on low power edge-devices. A typical implementation of this method involves multiple exit-branches at different stages of the forward pass of a large ML model, where each branch is composed of a smaller classifier~\cite{matsubara2022split}, with preset exit-logic. While this approach significantly reduces computation for some online samples, multiple exit-branches and complex exit-logic likely do not satisfy LHC trigger latency constraints~\cite{matsubara2022split}.

\section{PHAZE: Probabilistic Hashing And zkML-based Early-exit}
We present \texttt{PHAZE}~\cite{pratik_jawahar_2025_17370252}, a novel EE framework designed for a new generation of low latency ML-based triggers, built with the cryptographic primitives described in Sec.\ref{sec:intro}. For the LHC, feasibility studies for such a framework should broadly cover: computational constraints for online trigger systems; operational constraints laid by online hardware resources; physics performance in terms of trigger efficiency and other related parameters; verifiability for data quality monitoring and reproducing trigger decisions for downstream analyses.

\paragraph{Scope and assumptions:}
In this work we restrict ourselves to accelerating the inference process of a performant ML classifier that acts as the main triggering algorithm with the following assumptions: 
\begin{enumerate}
    \item Representative dataset: We assume that a large dataset corresponding to the input-space $\mathcal{I}$, is available to train on and serves as an acceptable representation of multiple classes of real events (backgrounds, noise and signals) commonly encountered at the LHC.
    \item Baseline model: Design choices for this large, high-accuracy model, $\mathbb{M}_\mathrm{full}$, like a single, large, general-purpose, physics foundation model versus multiple small specialized models; the model architecture; its training recipe etc., are left out of our scope.
    \item Early predictive sufficiency: We assume that early-layer activations, $\mathcal{A} \in \mathbb{R}^k$  of $\mathbb{M}_\mathrm{early} \subset \mathbb{M}_\mathrm{full}$, act as EE signals and are a strong proxy to the model decisions $\mathcal{D}=\{0,1\}^t$ (for $t$ different trigger classes), for a large enough subset $I \subseteq \mathcal{I}$. Here $\mathbb{R}^k$ is a $k$-dimensional real, latent activation vector space such that $\mathbb{M}_\mathrm{early}\colon I\rightarrow \mathbb{R}^k$. Design choices to fulfill this requirement may involve knowledge distillation (KD), fine-tuned adapter layers on top of a pre-trained $\mathbb{M}_\mathrm{full}$ or other such strategies and this choice is out of scope for our work.
    \item Injectivity of polynomial interpolation: We assume that a polynomial interpolation scheme that maps a vector $A^*\in \mathbb{R}^k$ to a polynomial $P_{A^*}(x)$ is available and injective. Here, $P_{A^*}(x)$ belongs to the space $\mathcal{P}_d(\mathbb{F}_p^m)$ of all $m$-variate polynomials of degree at most $d$.
\end{enumerate}

\subsection{Method}

\texttt{PHAZE} is designed to be a two-stage algorithm: a computation-intensive build phase that is not constrained by latency requirements or hardware choices and a low latency online phase that is expected to be a part of the online trigger algorithm.

\paragraph{Build phase:}
We start with a trained, large baseline model that gives the mapping $\mathbb{M}_\mathrm{full}\colon \mathcal{I}\rightarrow\mathcal{D}$. Based on the assumption above, this trained model can be approximated by the map $\mathrm{EE}\colon \mathcal{A}\rightarrow \mathcal{D}$, defined as the activation extraction step. While this EE can be statistically equivalent to KD, \texttt{PHAZE} does not smear over unseen events, and that is essential for discoveries. Each activation $A_j \in \mathcal{A}$ corresponding to $I_j \in \mathcal{I}$ is quantized to $A_j^* \in \mathbb{F}_p^k$. The choice between quantization during versus after training is left to experimental evaluation. We then generate an injective map for each $A_j^*$ to polynomial $P_{A_j^*}$ that passes through a predefined set of $k$ evaluation nodes, $\{x_1, ..., x_k\}\subset \mathbb{F}_p^m$ such that $P_{A_{j}^*}(x_i)=A_{j,i}^*$ for all $i \in \{1,...,k\}$. A rabin fingerprint is then generated as a hash for each such $P_{A_j^*}$ as described in Eq.\ref{eq:rabinfingerprint} with each polynomial evaluated at a constant, random challenge point $r=(r_1,...,r_m)$ sampled from $\mathbb{F}_p^m$. This fingerprint $h_j$ is then used as a key to populate a decision map , the Verifiable Decision Map (VDM), whose value is $D_j\in\mathcal{D}$. Next, we start the zk-STARK based proof generation, by reusing the polynomial interpolation and hashing steps. The computational integrity (CI) statement to be proven is formulated as a compound assertion $S_j$ for each $I_j$ given by:

\begin{quote}
    \textit{"The inference of $\mathbb{M}_\mathrm{full}$ on $I_j$ yields $D_j$} AND \textit{inference of $\mathbb{M}_\mathrm{full}$ on $I_j$ yields a quantized $A_j^*$, whose canonical polynomial $P_{A_j^*}$ evaluates to the hash value $h_j$ at the public challenge point $r$."}
\end{quote}
    
This statement is converted into an Algebraic Intermediate Representation (AIR), which consists of an execution trace of the whole system with corresponding polynomial constraints that hold if and only if the computations were executed correctly as explained in Sec.\ref{sec:zkML}. The zkML prover $\mathcal{P}$ then uses the AIR to generate a succinct proof $\pi_j$ to certify $S_j$. As discussed above, this proof generation, though computation-intensive, scales to reasonably large models, with the proofs being transparent for ad-hoc downstream auditing and verification. $\pi_j$ now certifies the validity of the decision map $\mathrm{VDM}\colon h_j\rightarrow D_j$. The proofs $\pi_j$ make the decision map resistant to corruptions such as accidental bit-flips and can then be archived.

\paragraph{Online phase:}
This phase is deployed at the trigger level and it starts with the same activation extraction step as in the build phase to produce $A_{\mathrm{new}}^*$ for each new event $I_\mathrm{new}$, ideally optimized for specialized hardware like FPGAs. Dedicated hardware cores then perform an on-the-fly (OTF) hashing step to directly evaluate $h_\mathrm{new}=P_{A_\mathrm{new}^*}(r)$ using the numerically stable Barycentric Lagrange interpolation and evaluation method~\cite{berrut2004barycentric} which skips a major portion of the polynomial definition in the build phase since we do not need to generate a new proof online. Then a fast VDM lookup is performed using $h_\mathrm{new}$ as the key to retrieve the trigger decision, $D'=\mathrm{VDM}[h_\mathrm{new}]$.

A map-miss occurs when no VDM key matches $h_\mathrm{new}$. This could indicate either a physics anomaly or an unknown detector anomaly. In either case, \texttt{PHAZE} inherently serves as an anomaly detector, with potential for incorporating specialized, downstream anomaly handling algorithms by buffering such events or by setting a default action (e.g., discard the event).

\begin{table}[h!]
\centering
\caption{Breakdown of computational tasks in the \texttt{PHAZE} framework.}
\label{tab:complexity}
\begin{tabular}{llll}
\hline
\textbf{Phase} & \textbf{Task} & \textbf{Latency per event} & \textbf{Computational complexity} \\
\hline
Build & Full Model Inference & $O(\text{ms - s})$ & $O(|\mathbb{M}_\mathrm{full}|)$ \\
 & EE Activation Extraction & $O(\mu\text{s})$ & $O(|\mathbb{M}_\mathrm{early}|)$ \\
 & Polynomial Interpolation & O(\text{ms - s}) & $O(\text{poly}(k \log k))$ \\
 & Probabilistic Hashing & $O(\mu\text{s})$ & $O(d)$ \\
 & zk-STARK Proof Generation & $O(\text{min})$ & $O(|\mathbb{M}_\mathrm{full}|\text{poly-log}(|\mathbb{M}_\mathrm{full}|))$ \\
 & ad-hoc Proof Verification & $O(\mu\text{s - ms})$ & $O(\text{poly-log}(|\mathbb{M}_\mathrm{full}|))$ \\
 & VDM Population & $O(\text{ns})$ & $O(1)$ per event \\
\hline
Online & \textbf{EE Activation Extraction} & \textbf{$O(\text{ns})$ (FPGA)} & \textbf{$O(|\mathbb{M}_\mathrm{early}|)$} \\
 & \textbf{On-the-Fly Hashing} & \textbf{$O(\text{ns})$ (FPGA)} & \textbf{$O(\sqrt{d})$ Erstin's method} \\
 & \textbf{VDM Lookup} & \textbf{$O(\text{ns})$ (FPGA)} & \textbf{$O(1)$ from BRAM} \\
\hline
\end{tabular}
\end{table}

\subsubsection{Latency estimates per event}
The net theoretical latency for the online trigger is thereby given by $T_\mathrm{online}=T_\mathrm{EE} + T_\text{OTF-hashing} + T_\text{VDM-lookup}$. A conservative estimate for $T_\mathrm{EE}$, for an optimized $\mathbb{M}_\mathrm{early}$ on an FPGA would be $100$-$200\mathrm{ns}$~\cite{govorkova2022lhc, odagiu2024ultrafast}. OTF-hashing using fast polynomial evaluation (e.g. Estrin's method~\cite{bodrato2011long}) takes as low as $\sim 20$ clock cycles on modern FPGAs, giving $T_\text{OTF-hashing}\approx 50\mathrm{ns}$, while a VDM lookup can be performed in a single clock cycle, giving $T_\text{VDM-lookup}\approx 2.5\mathrm{ns}$. Thereby a conservative estimate of $T_\mathrm{online}$ for a single event across a reasonable number of trigger classes would be in the range of $152.5$-$252.5\mathrm{ns}$ on modern FPGAs.

\subsubsection{Probabilistic guarantees and error components}
The model accuracy $\varepsilon(\mathbb{M}_\mathrm{full})$ is a major component of the net achievable trigger efficiency for \texttt{PHAZE}-like frameworks, and is framework-independent by assumption. The most prominent factor that leads to a drop in $\varepsilon(\mathbb{M}_\mathrm{full})$ is the probability of predictive insufficiency for the map $\mathrm{EE}\colon a_j \in \mathcal{A}\rightarrow d' \in \mathcal{D}$, given by $\Pr[d'\neq d_j]$, for ${d_j=\mathbb{M}_\mathrm{full}[I_j]}$. The operational bounds for this term depend on the EE strategy chosen and should be statistically estimated from experimental validation. The quantization of the latent activation vector can introduce prediction errors as well. However, this can be mitigated to an extent using representation learning techniques, like contrastive loss terms, that force the early layers to place different classes in distinct regions of the latent activation space. Events on the boundaries of these regions would have a higher chance of misclassification and thereby their contributions are encapsulated in the accuracy losses above. The hashing itself introduces an error when a collision occurs. However, for a $64$-bit hash and an expressive polynomial degree choice of $d=100$, the upper bound given by Eq.\ref{eq:schwartz-zippel} is $100/2^{64}$, which is computationally insignificant, but should be included in the net uncertainties regardless.

\subsubsection{Hardware resource feasibility}
The main accelerating factor that \texttt{PHAZE} takes advantage of, is moving most of the heavy computation to the build phase where we are not compute/latency constrained. For the online phase, evaluating $\mathrm{EE}\colon \mathcal{A}\rightarrow \mathcal{D}$ has shown to be feasible for reasonably large models using \texttt{hls4ml}. OTF-hashing using optimizations on Horner's rule or Estrin's method are viable on modern FPGAs (e.g., Xilinx Virtex 6)~\cite{xu2013efficient}. The VDM itself is a limiting component since its most efficient use requires storage in FPGA memory. For e.g., AMD UltraScale+ FPGA series provides $455\mathrm{Mb}$ memory. For a $64$-bit hash, each VDM entry would require say $72$ bits, for the hash and the decision across say $8$ trigger classes. This means a single FPGA can hold evaluations for $\frac{455\times10^6}{72}\approx 6.3\times10^6$ events. This size is not sufficient for a representative dataset and thereby strategies to enable fast distributed look-ups across multiple VDMs would be preferable. This is currently an open research question we look to address in our future work.

\section{Software package and Benchmarking results}
The \texttt{PHAZE}~\cite{pratik_jawahar_2025_17370252} software package (hosted on \texttt{GitHub}\footnote{Code available as the Zenodo-DOI for PHAZE in~\cite{pratik_jawahar_2025_17370252}}), is written mainly in \texttt{PYTHON} with custom \texttt{PYTHON}-\texttt{RUST} bindings to integrate contemporary cryptography packages generally written in \texttt{RUST}. \texttt{PHAZE} is designed for a wide range of benchmarking tasks on \texttt{PHAZE}-like trigger systems. The package (\texttt{v0.14.4}) contains the full build-phase pipeline. We are actively expanding towards full feasibility studies for LHC triggers.

\begin{figure}[ht!]
    \centering
    \includegraphics[width=0.45\linewidth]{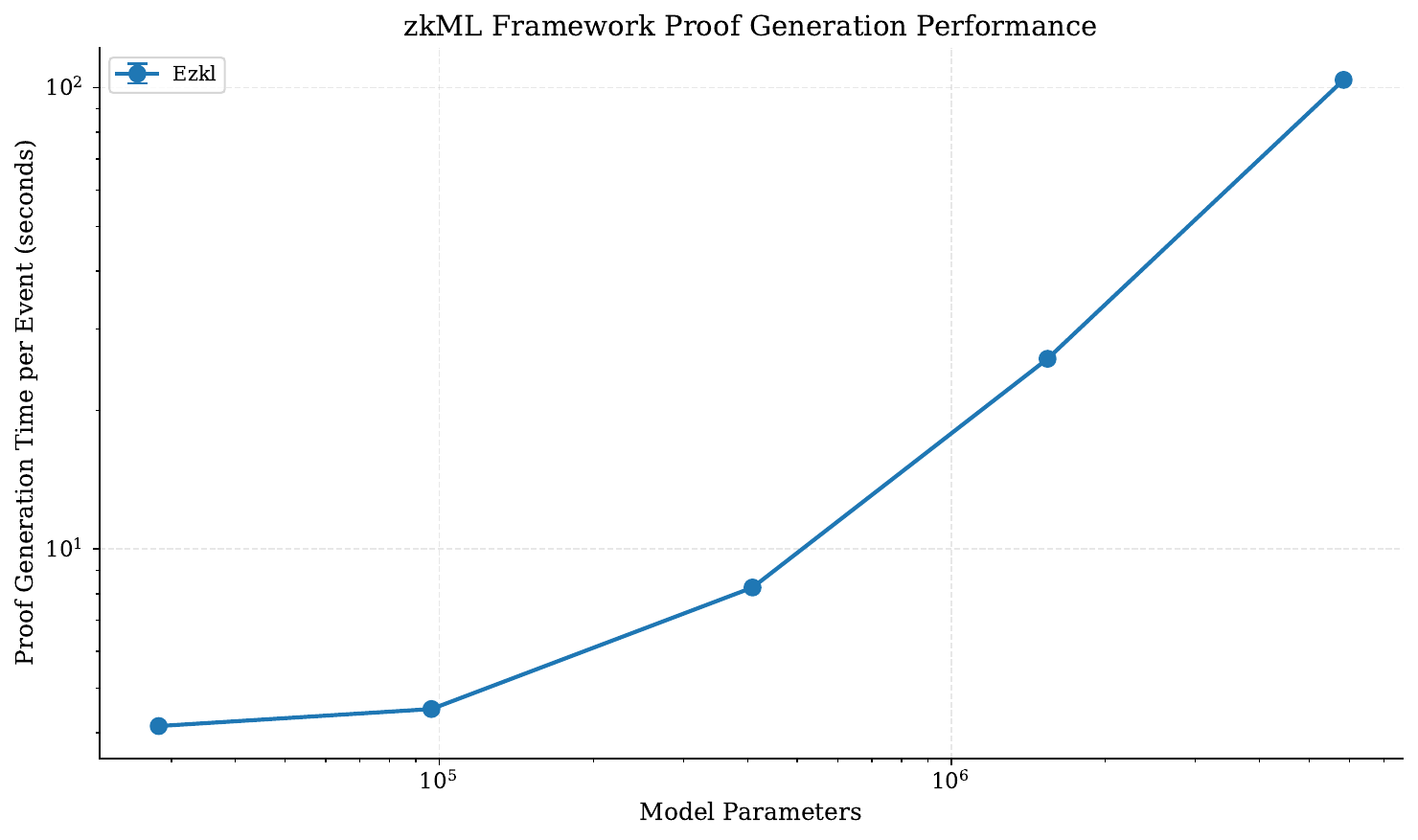}
    \includegraphics[width=0.45\linewidth]{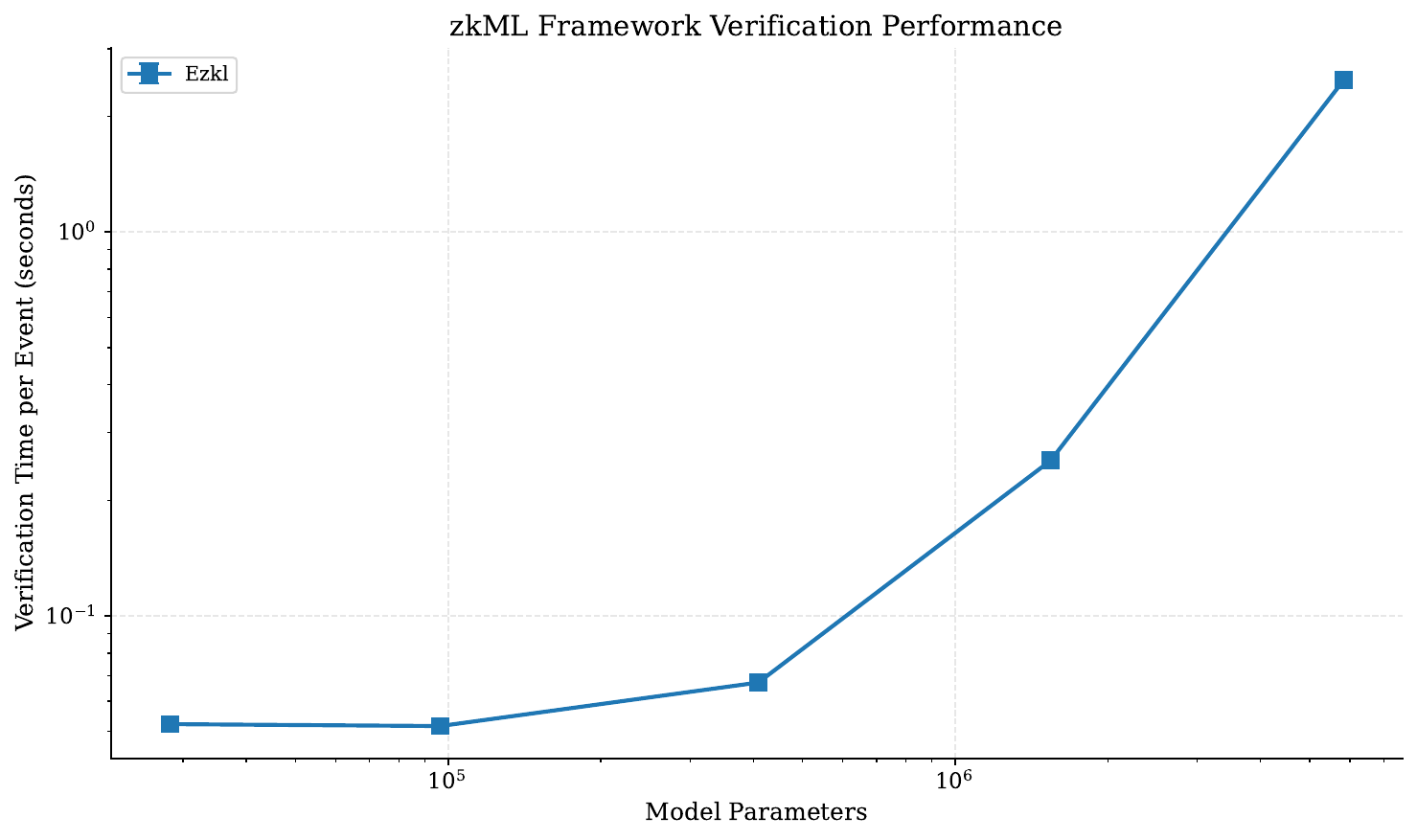}\\
    \includegraphics[width=0.45\linewidth]{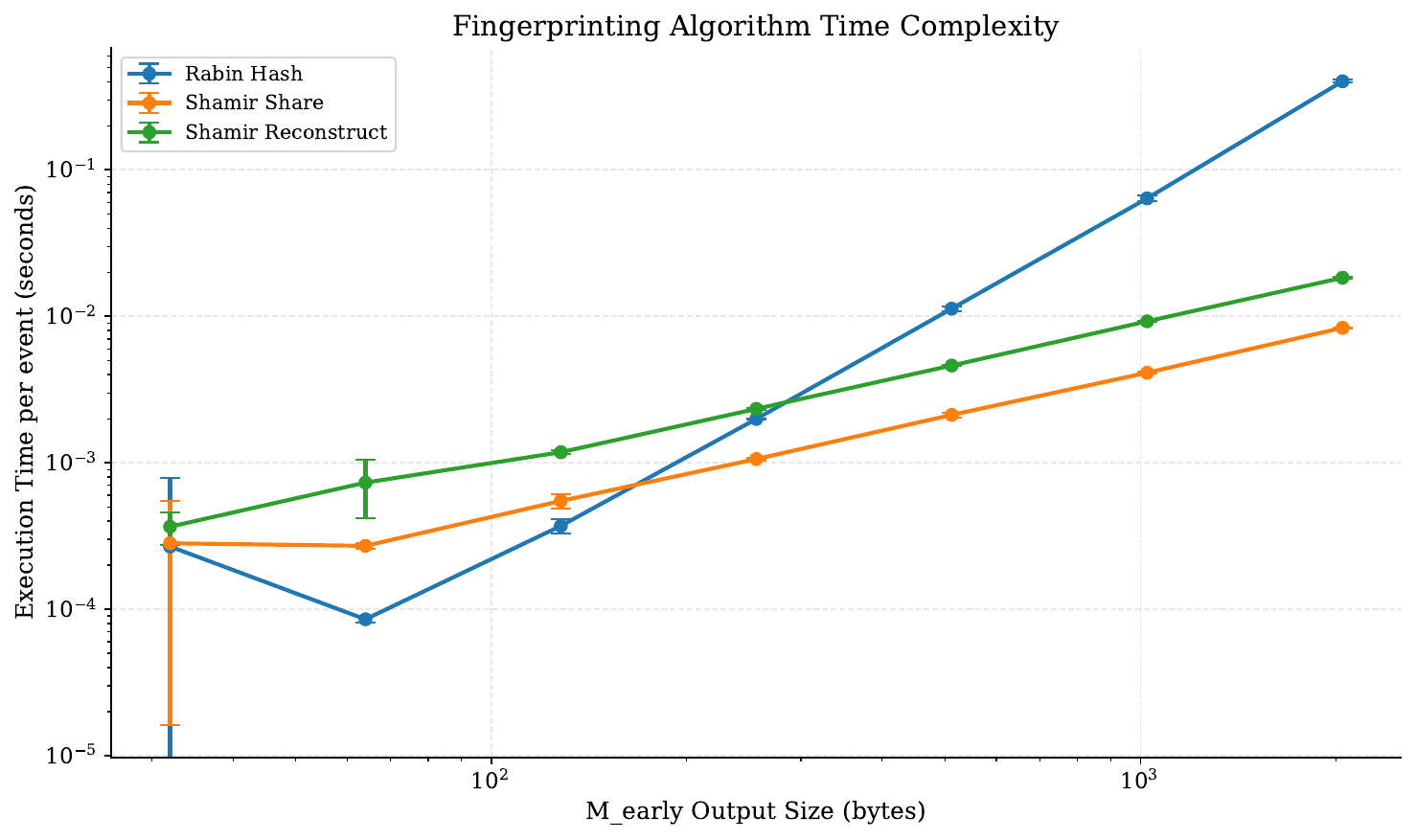}
    \includegraphics[width=0.45\linewidth]{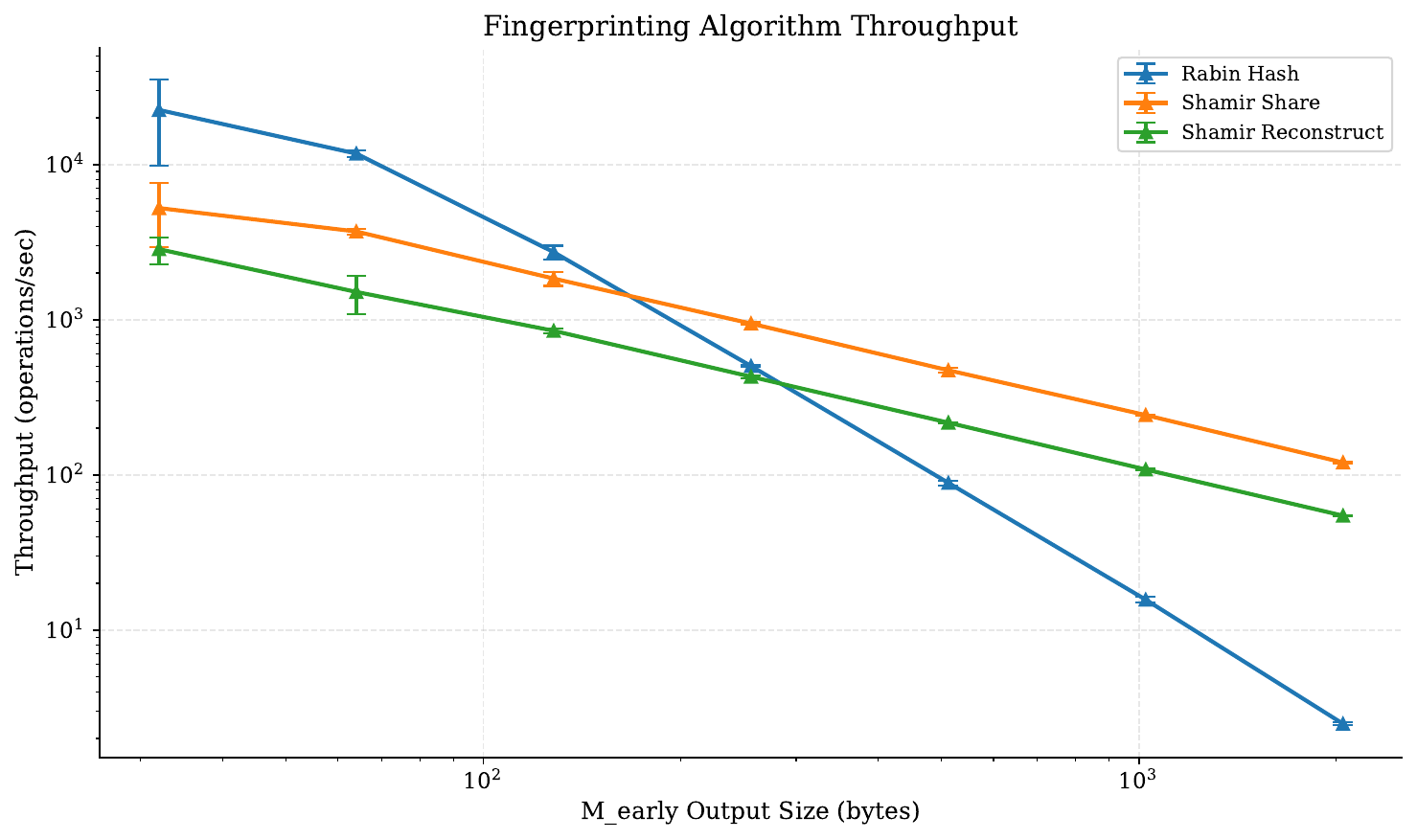}\\
    \includegraphics[width=0.45\linewidth]{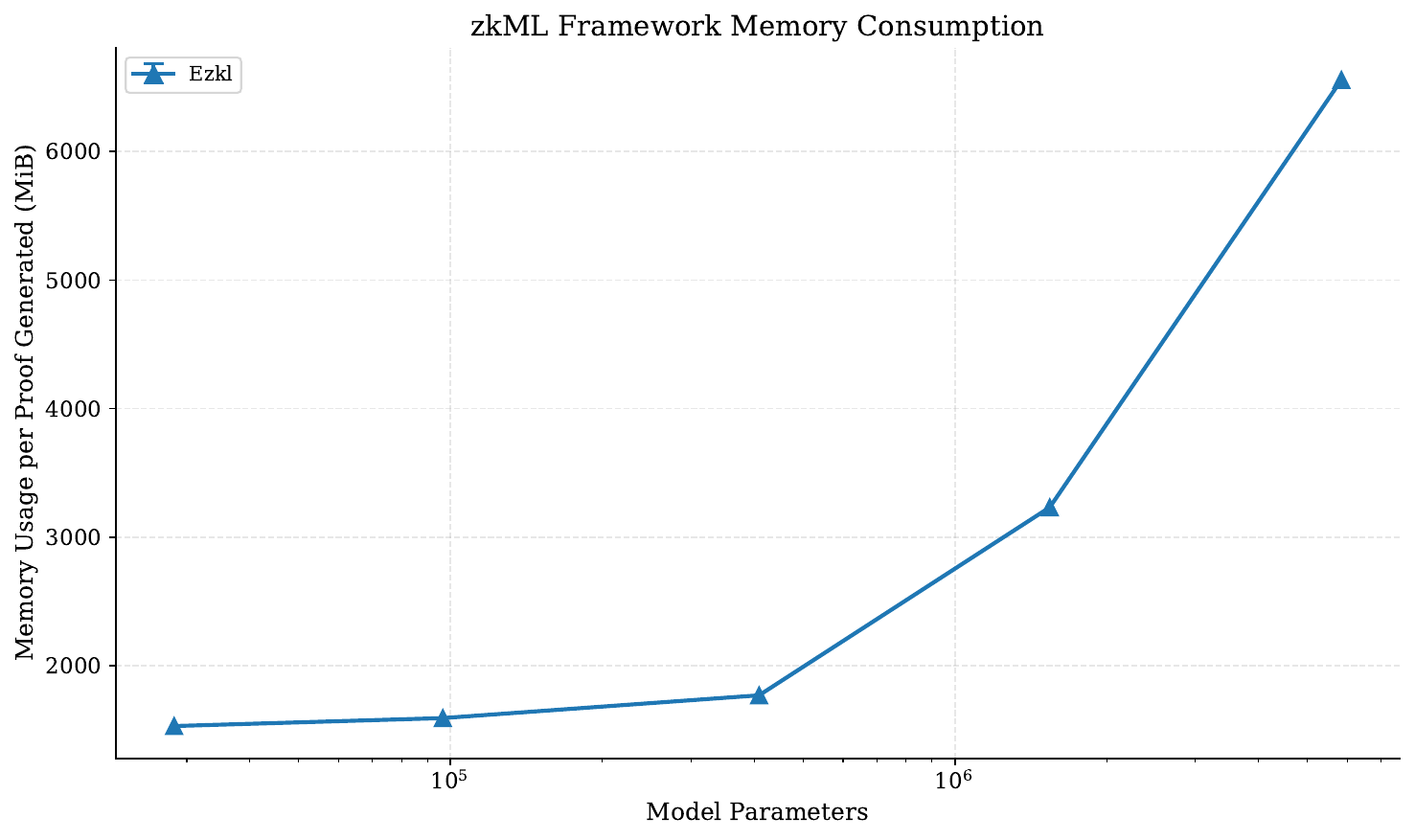}
    \includegraphics[width=0.45\linewidth]{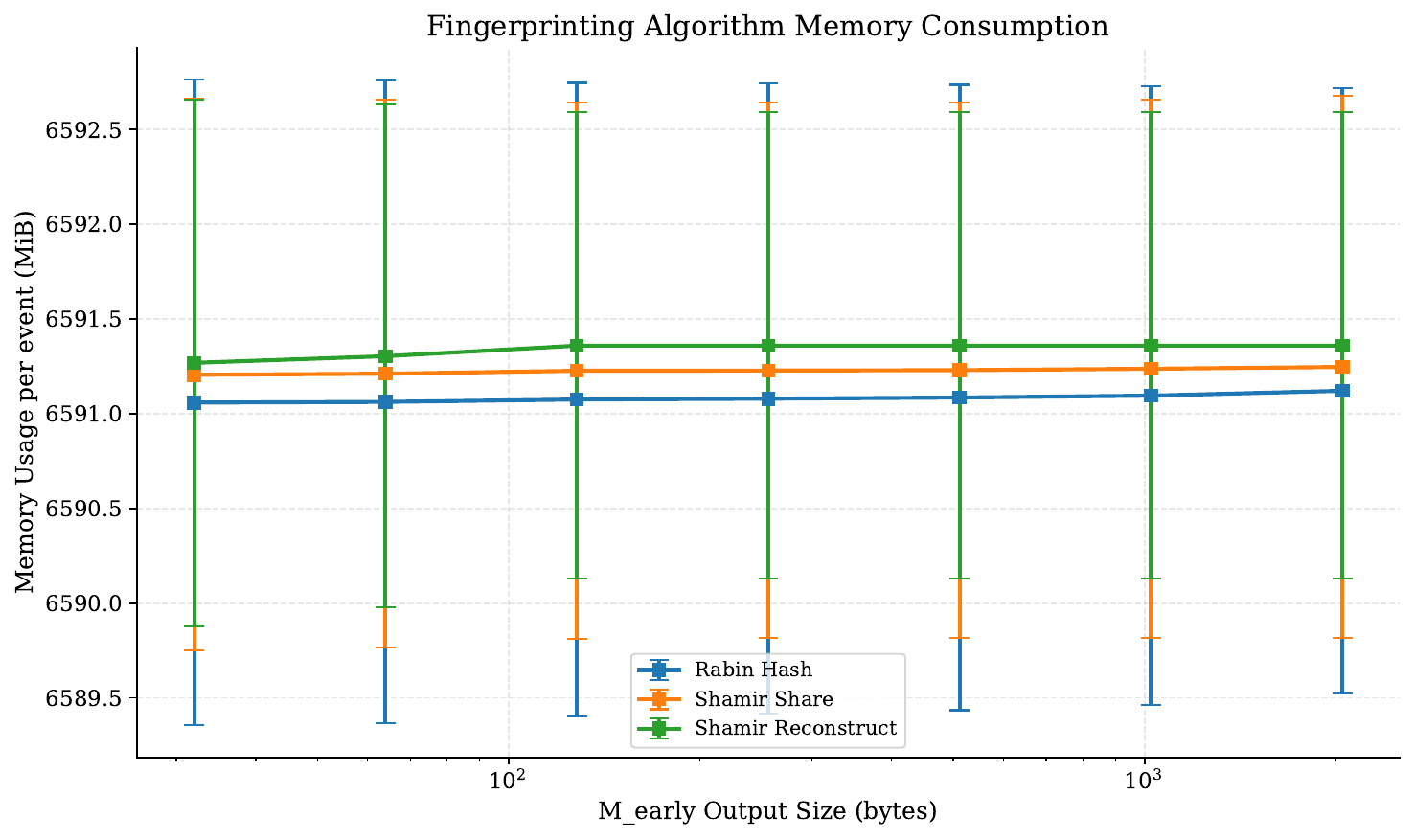}
    \caption{Early build-phase feasibility plots showing time complexity scaling for \texttt{ezkl} proof generation (top-left), verification (top-right) and fingerprinting (middle-left); net fingerprinting throughput (middle-right); peak memory consumption for \texttt{ezkl} proof generation (bottom-left) and fingerprinting (bottom-right). All values are reported per event, across $10$ independent trials. Note that for Shamir Secret Sharing to act as a fingerprinting algorithm, the timing information is roughly the sum of the Share and Reconstruct stages, while the peak memory consumption is the greater of the two at any given operation point.
    \label{fig:benchmarks}}
\end{figure}

\subsection{Early benchmark results: Build-scaling}
We construct $\mathbb{M}_\mathrm{full}$ as a DNN with $\sim7$M parameters and train it on the \texttt{MNIST} dataset~\cite{deng2012mnist}. We then extract multiple $\mathbb{M}_\mathrm{early}$ models at different parameter ratios from $5-95\%$ of $\mathbb{M}_\mathrm{full}$ parameter count. We use the \texttt{EZKL} package built on zk-SNARKs~\cite{chen2022review}. \texttt{EZKL} is chosen primarily for its simplicity for this early benchmarking study since it can generate proofs for any ML model exported to \texttt{ONNX}~\cite{onnxruntime}. It is to be noted that this is not the most computationally efficient zkML tool available, since tools that define ML models natively in \texttt{RUST} have been shown to be faster~\cite{moduluscostofintell}, but they come with the additional task of re-building \texttt{PYTHON}-based model architectures with custom layer-wise \texttt{RUST} code. We then compare Rabin fingerprinting and a competing algorithm Shamir Secret Sharing~\cite{shamir1979share}, to generate $64{\text -}\mathrm{bit}$ hashes of the different $M_{early}$ outputs. We run experiments on an Apple \texttt{MAC-M1 CPUs}. No optimizations have been attempted for these early benchmark results shown in Fig.\ref{fig:benchmarks}.

The key inferences we draw from Fig.\ref{fig:benchmarks} are connected to the areas of research that PHAZE-like triggers depend on. The time-complexity and memory scaling plots for the zkML stage show that for a relatively small $7$M-parameter model, proof generation per-event using an unoptimized zkML system takes $\sim10^2$ seconds. This points to a need for faster zkML systems with tools that allow easy conversion from common ML frameworks like \texttt{tensorflow}, \texttt{pytorch} to native operation in \texttt{RUST} or vice versa. Although fingerprinting is fast without many optimizations, there is scope for exploring other state-of-the-art strategies for better speed, and memory consumption while reducing collision probabilities and using intermediate polynomial conversions that can be seamlessly re-used at the zkML stage as well to improve net pipeline throughput. These are lines of research we intend to pursue as part of our future work and we encourage collaboration from the broader scientific community.

\section{Inherent capabilites and future directions}
\subsection{Map-miss anomaly detection}
Map-misses are an intrinsic tool that can provide "computationally free", low-level anomaly detection. Map-miss rates are strongly tied to representativeness of the dataset description, EE prediction sufficiency and VDM size and can be counter-productive if the above assumptions are not met. However, map-misses can strongly point to anomalous physics events or novel, unknown detector effects, that would otherwise be missed by current trigger algorithms. Such events would generally be discarded, or would require dedicated anomaly detection algorithms at the low-level trigger, which increases the computational overhead. Both these types of anomalies are essential to analyze for better long-term physics, as well as detector, performance. Leveraging map-misses in \texttt{PHAZE}-like trigger paradigms and combining it with powerful downstream trigger analyses or a dynamic VDM setup could potentially improve chances for discovery.

\subsection{Dynamic VDM}
The initial proposal envisions a static VDM, compiled once from a fixed dataset. However, the \texttt{PHAZE} framework readily supports dynamic updates, allowing the trigger system to adapt over time without requiring a full hardware redesign or resynthesis. The zkML prover-verifier setup is crucial to maintain the integrity of such a system.

\paragraph{Scheduled updates:}
The most direct approach involves periodically updating the VDM during planned LHC operational pauses, such as technical stops or shutdowns. During these periods, a new, more comprehensive VDM can be compiled offline. This updated map could incorporate knowledge gained from previous runs, adapt to changes in detector calibration and conditions, or, most significantly, include newly understood event topologies derived from classes of events with high map-miss rates.

\paragraph{Autonomous updates:}
A more ambitious form of a dynamic VDM would involve using knowledge accumulated online by downstream high-level triggers (HLT), and intermittently use beam downtime to update the VDM. This would require smaller baseline models that can simulate artificial maps between the event hashes and the HLT trigger decisions for event topologies with high map-miss rates. However, this would also require very slow online proof generation before the VDM can be altered. The current technology gap to realize a system like this is high, however, an autonomous, dynamic trigger system can greatly improve physics analyses, specially in the direction of new physics searches.

\section{Conclusions}
We present the \texttt{PHAZE} framework to potentially achieve nanosecond-order trigger latencies at the LHC using foundation model-scale ML algorithms. To the best of our knowledge, this is the first physics framework to utilize cryptographic primitives to enable fast inference and ensure verifiability of trigger algorithms, which is essential for a new era of dynamic, ML-based event selection at the LHC. We include comments on feasibility and the current assumptions are part of our ongoing efforts in realizing \texttt{PHAZE} as an experimental proof-of-concept system. \texttt{PHAZE} is also compatible as a building block for a larger distributed algorithm that is dynamically updated, while remaining verifiable, for flexible low level selections as a new paradigm for trigger design.

\bibliographystyle{JHEP.bst}
\bibliography{refs}

\end{document}